\documentclass[aps,prb,twocolumn,superscriptaddress,showpacs]{revtex4}

\usepackage{graphicx}
\usepackage{dcolumn}
\usepackage{bm}
\begin{document}


\title{Paramagnetism of the Co sublattice in ferromagnetic
Zn$_{1-x}$Co$_{x}$O films}

\author{A. Barla}
\altaffiliation{Now at: CELLS-ALBA, Edifici Cn.-M\'odul C/3 Campus
UAB, E-08193 Bellaterra, Barcelona, Spain, e-mail: abarla@cells.es}
\affiliation{Institut de Physique et Chimie des Mat\'eriaux de
Strasbourg, UMR 7504 ULP-CNRS, 23 rue du Loess, BP 43, F-67034
Strasbourg Cedex 2, France}
\author{G. Schmerber} \affiliation{Institut de Physique et
Chimie des Mat\'eriaux de Strasbourg, UMR 7504 ULP-CNRS, 23 rue du
Loess, BP 43, F-67034 Strasbourg Cedex 2, France}
\author{E. Beaurepaire}
\affiliation{Institut de Physique et Chimie des Mat\'eriaux de
Strasbourg, UMR 7504 ULP-CNRS, 23 rue du  Loess, BP 43, F-67034
Strasbourg Cedex 2, France}
\author{A. Dinia}
\affiliation{Institut de Physique et Chimie des Mat\'eriaux de
Strasbourg, UMR 7504 ULP-CNRS, 23 rue du  Loess, BP 43, F-67034
Strasbourg Cedex 2, France}
\author{H. Bieber}
\affiliation{Institut de Physique et Chimie des Mat\'eriaux de
Strasbourg, UMR 7504 ULP-CNRS, 23 rue du  Loess, BP 43, F-67034
Strasbourg Cedex 2, France}
\author{S. Colis}
\affiliation{Institut de Physique et Chimie des Mat\'eriaux de
Strasbourg, UMR 7504 ULP-CNRS, 23 rue du  Loess, BP 43, F-67034
Strasbourg Cedex 2, France}
\author{F. Scheurer}
\affiliation{Institut de Physique et Chimie des Mat\'eriaux de
Strasbourg, UMR 7504 ULP-CNRS, 23 rue du  Loess, BP 43, F-67034
Strasbourg Cedex 2, France}
\author{J.-P. Kappler}
\affiliation{Institut de Physique et Chimie des Mat\'eriaux de
Strasbourg, UMR 7504 ULP-CNRS, 23 rue du  Loess, BP 43, F-67034
Strasbourg Cedex 2, France}
\author{P. Imperia}
\affiliation{Hahn Meitner Institut, Glienicker Strasse 100,
D-14109 Berlin, Germany}
\author{F. Nolting}
\affiliation{Swiss Light Source, Paul Scherrer Institut, CH-5232
Villigen PSI, Switzerland}
\author{F. Wilhelm}
\affiliation{European Synchrotron Radiation Facility, B.P. 220,
F-38043 Grenoble Cedex 9, France}
\author{A. Rogalev}
\affiliation{European Synchrotron Radiation Facility, B.P. 220,
F-38043 Grenoble Cedex 9, France}
\author{D. M\"uller}
\affiliation{InESS, UMR 7163 CNRS, 23 rue du  Loess, BP 20CR,
F-67037 Strasbourg Cedex 2, France}
\author{J. J. Grob}
\affiliation{InESS, UMR 7163 CNRS, 23 rue du  Loess, BP 20CR,
F-67037 Strasbourg Cedex 2, France}

\date{\today}

\begin{abstract}
Using the spectroscopies based upon x-ray absorption, we have
studied the structural and magnetic properties of
Zn$_{1-x}$Co$_{x}$O films ($x$\,=\,0.1 and 0.25) produced by
reactive magnetron sputtering. These films show ferromagnetism with
a Curie temperature $T_{\mathrm{C}}$ above room temperature in bulk
magnetization measurements. Our results show that the Co atoms are
in a divalent state and in tetrahedral coordination, thus
substituting Zn in the wurtzite-type structure of ZnO. However,
x-ray magnetic circular dichroism at the Co \textit{L}$_{2,3}$ edges
reveals that the Co 3\textit{d} sublattice is paramagnetic at all
temperatures down to 2~K, both at the surface and in the bulk of the
films. The Co 3\textit{d} magnetic moment at room temperature is
considerably smaller than that inferred from bulk magnetisation
measurements, suggesting that the Co 3\textit{d} electrons are not
directly at the origin of the observed ferromagnetism.
\end{abstract}


\maketitle


Among the most investigated topics in the field of spin electronics,
dilute magnetic semiconductors (DMSs) occupy a prominent position,
because they would allow one to exploit efficiently the spin and the
charge of the electrons in the same device. In fact electronic
devices have been working for decades omitting the spin of the
electron. In 1990 Datta and Das proposed a new magneto-electronic
device (a field effect transistor),\cite{DaD90} whose practical
realisation has been hindered by the weak spin injection efficiency
from a ferromagnet to a semiconductor. A ferromagnetic semiconductor
would constitute therefore an alternative route towards the
efficient spin injection into normal semiconductors. Up to very
recently, however, the main concern was related to the low Curie
temperature of the known DMSs, which is well below room temperature
and precludes therefore potential applications.\cite{EWC02} A
significant breakthrough was achieved recently, since room
temperature ferromagnetism was predicted \cite{DOM00} and observed
\cite{FTY05,VFL04,KNG05,SHK06,TOG02,And03} for semiconductors such
as GaN and ZnO doped with Co, Mn or other transition metals.
However, many reports remain controversial and the nature of the
magnetic coupling has not been revealed yet. In fact, the original
prediction of high $T_{\mathrm{C}}$ ferromagnetism in these systems
by Dietl et al. \cite{DOM00} lies on the assumption that they can be
properly doped with $p$-type carriers, which would mediate the
magnetic interactions. However, in order to account for the numerous
experimental observations of ferromagnetism in $n$-type ZnO,
alternative models have been proposed, which are mainly relying on
the presence of defects (like for example vacancies or
interstitials).\cite{CVF05,PaC05} In most of these models, the
presence of a magnetic impurity such as Co or Mn is a necessary
ingredient for the appearance of ferromagnetism, but other models
show that this might not be needed and that ferromagnetism can
appear also in undoped oxides.\cite{EYS02} To date, however, there
has not been any clear experimental proof of the validity of any of
these models and of the role of the magnetic dopants.

In order to tackle these problems, we have performed extensive
studies by x-ray absorption spectroscopies of Co doped ZnO films,
which are ferromagnetic
above room temperature according to bulk magnetization studies.
Our results show that, within the sensitivity limits
of these techniques, Co substitutes Zn in the wurtzite structure
which is typical for ZnO and that the Co magnetic sublattice is
paramagnetic, with strong antiferromagnetic correlations.

Films of Zn$_{1-x}$Co$_{x}$O ($x$\,=\,0.1 and 0.25) were grown on
Al$_{2}$O$_{3}$(0001) substrates by reactive magnetron
co-sputtering, using pure Zn and Co targets. The working pressure
was a mixture of argon at 5$\cdot$10$^{-3}$\,Torr and oxygen at
1.5$\cdot$10$^{-3}$\,Torr. The thickness of the films was fixed at
100~nm and their composition was controlled by adjusting the
sputtering power applied to the Co and Zn targets. During the
deposition, the substrates were kept at 600\,$^{\circ}$C. The films
are transparent and standard x-ray diffraction experiments reveal
that they are highly textured along the \textit{c}-axis of the
hexagonal wurtzite structure (space group
\textit{P}6$_{3}$\textit{mc}). The Zn$_{0.9}$Co$_{0.1}$O film was
additionally implanted with 0.5\%~As$^{+}$ ions in order to increase
the number of free carriers. While Zn$_{0.75}$Co$_{0.25}$O is
clearly insulating (resistance of the order of M$\Omega$ at room
temperature), As implantation induces a considerable reduction of
the electrical resistance at room temperature and Hall effect
measurements indicate that the Zn$_{0.9}$Co$_{0.1}$O:As film is
$n$-doped with a carrier concentration of
$\sim$2$\cdot$10$^{19}$~cm$^{-3}$ (Ref.~\onlinecite{NCS06}). The
optical absorption spectra, as measured by UV-Vis transmission
spectroscopy at room temperature, show the absorption bands which
are characteristic of \textit{d}-\textit{d} transitions in
tetrahedrally coordinated high spin Co$^{2+}$ (at wavelengths
$\lambda$~$\sim$~570, 615 and 655~nm), thus suggesting that Co
substitutes for Zn in the wurtzite structure of ZnO
(refs.~\onlinecite{DSM05,NCS06}). Moreover, a clear shift of the
absorption edge towards higher wavelengths is observed as the Co
concentration increases. The structural, optical and electrical
properties of these films have already been discussed in detail in
refs.~\onlinecite{DSM05,NCS06}.

\begin{figure}[t]
\includegraphics[scale=0.8,clip=]{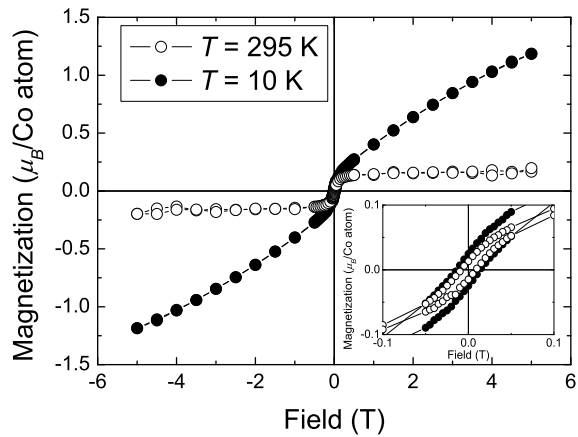}
\caption{\label{fig:hystsquid} Magnetization curves measured by
SQUID magnetometry at different temperatures on
Zn$_{0.9}$Co$_{0.1}$O:As thin films, after subtraction of the
diamagnetic contribution of the Al$_{2}$O$_{3}$ substrate. The
magnetic field was applied perpendicular to the crystallographic
\textit{c} axis (i.~e. parallel to the film's surface). The inset
shows a zoom of the region of low magnetic fields that evidences the
presence of a hysteresis.}
\end{figure}
The bulk magnetic properties of the Zn$_{1-x}$Co$_{x}$O films were
studied by superconducting quantum interference device (SQUID)
magnetometry in the temperature range between 5 and 295~K in
magnetic fields up to 5~T. Figure~\ref{fig:hystsquid} shows the
magnetization curves of Zn$_{0.9}$Co$_{0.1}$O:As measured at room
temperature and at 10~K as a function of applied magnetic field,
whose direction was perpendicular to the crystallographic
\textit{c}-axis of the film. At 295~K, a hysteresis loop opens at
small fields, with a coercive field of the order of 9~mT, and the
magnetization shows little dependence on the magnetic field for
fields larger than 0.5~T. These findings indicate that the film is
ferromagnetic, with a Curie temperature higher than room
temperature. The magnetization measured in a field of 10~mT shows
little dependence on temperature between 295 and 50~K. However,
below 50~K the susceptibility increases sharply with decreasing
temperature, thus indicating the presence of paramagnetic moments.
\cite{NCS06} This is confirmed by the field dependence of the
magnetization measured at 10~K (see figure~\ref{fig:hystsquid}),
which is the superposition of a ferromagnetic (coercive field of
$\approx$14~mT) and a paramagnetic component. Similar results were
obtained on the Zn$_{0.75}$Co$_{0.25}$O film (see, for example,
ref.~\onlinecite{DSM05}).

\begin{figure}[t]
\includegraphics[scale=0.8,clip=]{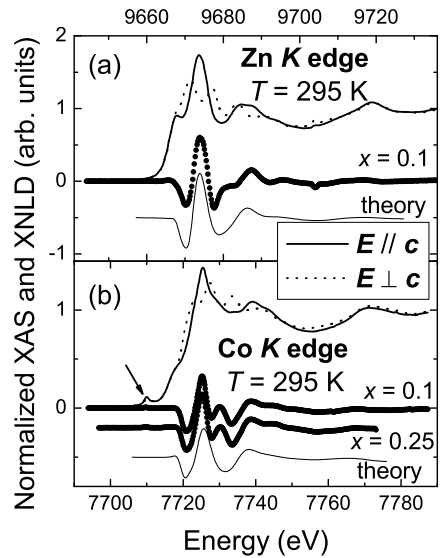}
\caption{\label{fig:xnlskedge} The room temperature XAS spectra of
Zn$_{0.9}$Co$_{0.1}$O:As and Zn$_{0.75}$Co$_{0.25}$O as measured at
(a) the Zn and (b) the Co \textit{K} edge, with the polarization
vector of the x-rays parallel (thick solid line) and perpendicular
(thick dotted line) to the crystallographic \textit{c}-axis, and
their difference (XNLD, full circles). The calculated XNLD is shown
as a thin solid line and is shifted vertically for clarity as well
as the XNLD curve for Zn$_{0.75}$Co$_{0.25}$O at the Co $K$ edge.}
\end{figure}
Information about the structural and electronic properties of
Zn$_{1-x}$Co$_{x}$O was obtained by x-ray absorption spectroscopy
(XAS) and x-ray natural linear dichroism (XNLD) at the \textit{K}
edges of Zn and Co (1\textit{s}$\rightarrow$4\textit{p}
transitions), performed at beamline ID12 of the European Synchrotron
Radiation Facility, Grenoble, France, at room temperature and in
total fluorescence yield mode. The measurements were done by
rotating the direction of the polarization vector
\textit{\textbf{E}} with respect to the crystallographic
\textit{c}-axis. Figures~\ref{fig:xnlskedge}(a) and (b) show the
polarization dependent XAS spectra of Zn$_{0.9}$Co$_{0.1}$O:As and
Zn$_{0.75}$Co$_{0.25}$O at the Zn and Co \textit{K} edges. Identical
results were obtained on other films with Co concentrations
0~$\leq$~$x$~$\leq$~0.25. The strong anisotropy of the x-ray
absorption at both \textit{K} edges, due to the preferential growth
of the films with the \textit{c}-axis perpendicular to the surface,
leads to the observation of a strong XNLD signal. The \textit{K}
edge XAS spectra of Zn$_{1-x}$Co$_{x}$O films have been calculated
by using the \textit{ab-initio} code FDMNES,\cite{Jol01} in the mode
which uses the multiple scattering formalism on a muffin-tin
potential. Figures~\ref{fig:xnlskedge}(a) and (b) show a comparison
of the experimental XNLD signals with those extracted from the
\textit{ab-initio} calculations on a 77-atom cluster (6~\AA~radius).
The Zn \textit{K} edge XNLD can well be approximated by that of pure
ZnO (as shown in figure~\ref{fig:xnlskedge}(a)), thus confirming
that the introduction of Co does not significantly change the
structural properties of the ZnO matrix. Some small discrepancies
between experiment and model are evident in the region just above
the absorption edge and are due to the limited size of the cluster
used in the calculations (which was chosen as a good compromise
between accuracy of the calculation and computational time). The Co
\textit{K} edge XNLD has been calculated by artificially replacing
all the Zn$^{2+}$ ions by Co$^{2+}$ ions in the wurtzite-type
structure of ZnO (without any change in the lattice constants).
Although this is an approximation, it can be considered as justified
by the fact that the Co $K$ edge XAS and XNLD spectra are the same
for the whole range of Co concentrations between 5 and 25\%. Even in
this case the agreement between experiment and calculation is very
good (see figure~\ref{fig:xnlskedge}(b)), thus confirming that Co is
occupying substitutional positions. It is especially important to
notice that the calculated XNLD signal has not been rescaled with
respect to the experimental one and this suggests that all Co atoms
occupy positions with the same symmetry in the lattice. A small
amount ($\leq$5\% of the total Co concentration) of clusters of
metallic Co, for example, would not only have a visible influence on
the shape of the XAS spectrum, but it would also reduce the measured
XNLD amplitude (to which metallic Co does not contribute) with
respect to the calculated one.

\begin{figure}[t]
\includegraphics[scale=0.8,clip=]{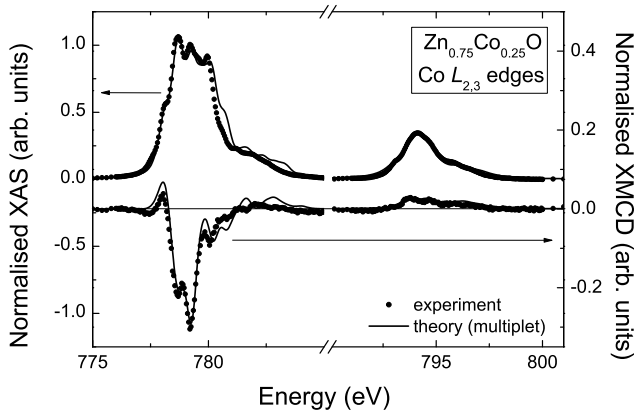}
\caption{\label{fig:xmcd} Isotropic XAS and XMCD spectra measured
(full circles, TEY) and calculated within the theory of atomic
multiplets (thick solid lines) for Zn$_{0.75}$Co$_{0.25}$O at the Co
\textit{L}$_{2,3}$ edges.}
\end{figure}
In order to investigate independently the magnetic properties of the
Co sublattice, XAS and x-ray magnetic circular dichroism (XMCD)
measurements at the Co \textit{L}$_{2,3}$ edges
(2\textit{p}$\rightarrow$3\textit{d} transitions) were performed at
beamlines UE56/2 and UE46 of BESSY, Berlin, Germany
(Zn$_{0.9}$Co$_{0.1}$O:As film) and SIM of the Swiss Light Source,
Villigen, Switzerland (Zn$_{0.75}$Co$_{0.25}$O film). The XAS signal
was detected simultaneously in both total electron yield (TEY) and
total fluorescence yield (TFY) modes, ensuring both surface and bulk
sensitivity, respectively. The XAS and XMCD spectra measured at the
Co \textit{L}$_{2,3}$ edges in TEY mode show a pronounced multiplet
structure at both edges (see Fig.~\ref{fig:xmcd} for the case of
$x$~=~0.25), which is typical for cobalt in a non-metallic
environment. These spectra are very similar to those reported for
Zn$_{1-x}$Co$_{x}$O films in ref.~\onlinecite{KIH05}. In order to
simulate both XAS and XMCD spectra, multiplet calculations were
performed with a program based on Cowan's Hartree-Fock atomic code
with point charge crystal field.\cite{Cow81} Figure~\ref{fig:xmcd}
shows that an excellent agreement between experiment and theory can
be achieved if we consider Co$^{2+}$ ions (3\textit{d}$^{7}$
configuration) occupying sites with \textit{C}$_{3v}$ point
symmetry, as expected if Co substitutes Zn. The crystal field
parameters which best fit simultaneously both XAS and XMCD spectra
are 10\textit{Dq}~=~-0.47~eV, \textit{D$\sigma$}~=~0.06~eV and
\textit{D$\tau$}~=~-0.03~eV.

\begin{figure}[t]
\includegraphics[scale=0.8,clip=]{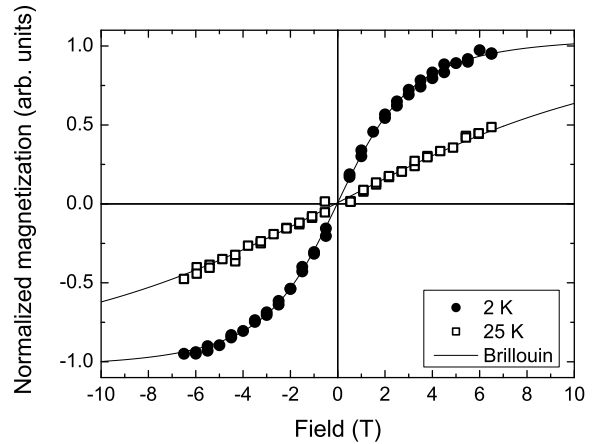}
\caption{\label{fig:hystxmcd} Normalized magnetization curves of
Zn$_{0.9}$Co$_{0.1}$O:As measured at the Co \textit{L}$_{3}$ edge in
TEY at $T$~=~2~K (full cicrcles) and at $T$~=~25~K (empty squares).
The magnetization curves are compared with Brillouin functions
calculated for $S$~=~3/2 and $L$/$S$~=~0.7 (solid lines).}
\end{figure}
Element selective magnetization curves have been recorded in both
TEY and TFY at 2, 5 and 25~K on Zn$_{0.9}$Co$_{0.1}$O:As, by
scanning the magnetic field while keeping the incident energy fixed
at the maximum of the XMCD signal at the Co \textit{L}$_{3}$ edge
($E$~=~779.2~eV), at both normal and 45$^{\circ}$ incidence.
Figure~\ref{fig:hystxmcd} shows as an example the curves measured in
TEY at 2 and 25~K at 45$^{\circ}$ incidence, which are normalized to
a saturation magnetization of 1. Although these measurements do not
give directly an absolute value of the Co magnetic moment as a
function of the applied magnetic field, they give a first clear
indication about the field dependence of the magnetization of Co in
Zn$_{0.9}$Co$_{0.1}$O:As. The curves measured at all temperatures
can well be accounted for by Brillouin functions (shown as
continuous lines in the figure), calculated with the fixed values of
$S$~=~3/2 (for Co$^{2+}$) and $L$/$S$~=~0.7 (this value of $L$/$S$
is determined through the application of the magneto-optical sum
rules to the XAS and XMCD spectra as discussed below). Strikingly
and unexpectedly, the results displayed in figure~\ref{fig:hystxmcd}
show therefore the presence of a purely paramagnetic contribution of
the Co 3\textit{d} sublattice to the total magnetization of the
Zn$_{0.9}$Co$_{0.1}$O:As film. This paramagnetic behaviour is
confirmed by the absence of any detectable hysteresis in the
magnetization curves measured by XMCD, both at normal and
45$^{\circ}$ incidence. Similar results were obtained in TFY mode,
so that we cannot observe any difference in the magnetic properties
of the surface (measurements in TEY mode) and of the bulk
(measurements in TFY mode), as well as on the
Zn$_{0.75}$Co$_{0.25}$O film.

More quantitative information can be obtained by the application of
the magneto-optical sum rules, which allow one to evaluate
separately both the spin and the orbital magnetic moments carried by
the Co atoms.\cite{TCS92,CTA93} We suppose a pure 3\textit{d}$^{7}$
configuration for the Co atoms (in agreement with the multiplet and
the \textit{ab-initio} calculations of the XAS spectra), which fixes
the number of holes in the 3\textit{d} shell to three. In this case
we obtain for Zn$_{0.9}$Co$_{0.1}$O:As a spin magnetic moment
$m_{S}$~=~0.81(8)~$\mu_{\mathrm{B}}$ and an orbital magnetic moment
$m_{L}$~=~0.27(3)~$\mu_{\mathrm{B}}$ at $T$~=~2~K and $H$~=~6.5~T in
normal incidence and TEY. Slightly larger values
[$m_{S}$~=~0.97(10)~$\mu_{\mathrm{B}}$ and
$m_{L}$~=~0.31(3)~$\mu_{\mathrm{B}}$] are obtained at 45$^{\circ}$
incidence (TEY), suggesting the presence of a small anisotropy of
the 3\textit{d} magnetic moment. However, the ratio $m_{L}$/$m_{S}$
keeps a constant value of $\sim$0.35(4). The total magnetic moment
$m_{\mathrm{tot}}$~$\approx$~1.2(1)~$\mu_{\mathrm{B}}$ is
considerably lower than the value of $\sim$~4~$\mu_{\mathrm{B}}$,
which would be expected for $S$~=~3/2 and for the value
$L$/$S$~=~2$m_{L}$/$m_{S}$~$\sim$~0.7 determined experimentally from
the application of the sum rules. Similar results were obtained on
Zn$_{0.75}$Co$_{0.25}$O, but with an even more reduced magnetic
moment of $m_{\mathrm{tot}}$~=~0.55(6)~$\mu_{\mathrm{B}}$/Co atom at
5~K and 5~T, measured in TEY mode and normal incidence. This
reduction of the total magnetization has already been observed in
most bulk Zn$_{1-x}$Co$_{x}$O samples with large Co concentrations
and has been attributed to an inhomogeneous distribution of the Co
atoms in the ZnO lattice.\cite{BVC05,RSC03} This gives rise to
strong local Co-Co antiferromagnetic correlations, mediated by
neighboring O atoms, and can possibly lead to the onset of
antiferromagnetic order in the Co rich regions, which would
therefore not contribute to the total magnetization. We can
tentatively ascribe to this mechanism the observation of a low
saturation magnetization of the Co sublattice in our samples as well
as the lower total magnetic moment of Zn$_{0.75}$Co$_{0.25}$O as
compared to Zn$_{0.9}$Co$_{0.1}$O:As.

The total Co magnetic moment in Zn$_{0.9}$Co$_{0.1}$O:As decreases
rapidly with increasing temperature
[$m_{\mathrm{tot}}$~=~0.65(7)~$\mu_{\mathrm{B}}$ at $T$~=~25~K and
$H$~=~6.5~T] and almost vanishes at room temperature, where
$m_{\mathrm{tot}}$~=~0.05(3)~$\mu_{\mathrm{B}}$ at $H$~=~4~T. This
is in stark contrast with the result of the bulk magnetization
measurements, which show a ferromagnetic signal at least four times
larger under the same conditions (see fig.~\ref{fig:hystsquid}). The
XAS and XMCD spectra measured in TFY mode are qualitatively similar,
but the strong self-absorption in the sample alters significantly
the branching ratio between the \textit{L}$_{2}$ and
\textit{L}$_{3}$ absorption edges, thus making the application of
the sum-rules meaningless. However, they qualitatively confirm the
strong decrease of the total Co magnetic moment with increasing
temperature also in the bulk of the sample.

All these findings are in stark contrast with the results of the
bulk magnetization measurements and strongly support the idea that
the Co doping is not the primary cause of the high temperature
ferromagnetism observed in the Zn$_{1-x}$Co$_{x}$O films. The
paramagnetic contribution to the bulk magnetization, as obtained by
subtracting the magnetization measured at 295~K from that at 10~K,
amounts to $m_{\mathrm{par}}$~=~1.0(1)~$\mu_{\mathrm{B}}$/Co atom at
5~T and 10~K for Zn$_{0.9}$Co$_{0.1}$O:As and coincides therefore,
within the error, with the total Co 3\textit{d} magnetic moment
determined by the XMCD measurements. These findings allow us to
attribute all the paramagnetic signal to the 3\textit{d} moments of
the Co sublattice and, at the same time, to exclude their
significant contribution to the ferromagnetism. We cannot, however,
exclude the presence of a tiny ferromagnetic Co 3\textit{d} moment
(as expected in the case of long range ferromagnetic order), below
the detection limit of our XMCD measurements
($\sim$0.01$\mu_{\mathrm{B}}$), but this would be much lower than
the ferromagnetic moment measured by SQUID magnetometry at room
temperature.

\begin{figure}[t]
\includegraphics[scale=1.0,clip=]{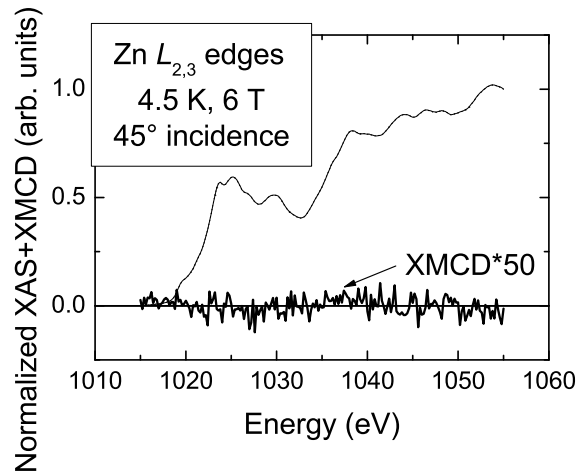}
\caption{\label{fig:znledge} The XAS spectra of
Zn$_{0.9}$Co$_{0.1}$O:As measured at the Zn \textit{L}$_{2,3}$ edges
in TFY mode, 45$^{\circ}$ incidence, $T$~=~4.5~K and $H$~=~6~T, with
right circularly polarized (thin solid line) and left circularly
polarized (thin dotted line) x-rays, and the corresponding XMCD
spectrum (thick solid line).}
\end{figure}
In order to further investigate the origin of ferromagnetism in our
Zn$_{0.9}$Co$_{0.1}$O:As film, we have looked for the possible
magnetic polarization of the Zn sublattice. The electronic
configuration for Zn$^{2+}$ is formally 3\textit{d}$^{10}$, so that
a 3\textit{d} magnetic moment cannot be expected to be present.
However, the presence of vacancies and/or interstitial Zn atoms
could be at the origin of an unfilled 3\textit{d} shell, which could
in turn carry a magnetic moment. Figure~\ref{fig:znledge} shows the
XAS and XMCD spectra measured in TFY at the Zn \textit{L}$_{2,3}$
edges at $T$~=~4.5~K and $H$~=~6~T. The XAS spectra do not show the
presence of any strong white line, as expected for a fairly pure
3\textit{d}$^{10}$ configuration. There is no evidence of an XMCD
signal in the measured energy region, down to the noise level of
0.15\% of the jump at the edge. This excludes the presence of any
detectable ferromagnetic polarization of the Zn \textit{s} and
\textit{d} shells.

These results indicate therefore that the 3\textit{d} electronic
shells of the cations in our Zn$_{1-x}$Co$_{x}$O films do not carry
any measurable ferromagnetic moment, contrary to what is usually
assumed in the theoretical models proposed to account for the
unexpected magnetic properties of this and related systems. This
seems to suggest that mostly the anion sublattice (i.~e. the oxygen
ions) might be responsible for the ferromagnetic moment observed in
Zn$_{1-x}$Co$_{x}$O (for example through the presence of vacancies
or interstitials) and that this and other doped oxides might exhibit
properties similar to those of pure HfO$_{2}$
(ref.~\onlinecite{VFC04}) or TiO$_{2}$ (ref.~\onlinecite{HSP06}).
Future investigations should therefore also be directed towards
obtaining reliable XMCD signals at the O \textit{K} edge in such
systems.

In summary, we have investigated the structural and magnetic
properties of films of Zn$_{1-x}$Co$_{x}$O produced by reactive
magnetron co-sputtering. They show ferromagnetism with a Curie
temperature $T_{\mathrm{C}}$ above room temperature in bulk
magnetization measurements. At temperatures below $\sim$50~K a clear
paramagnetic component appears, which dominates at the lowest
temperatures. Our x-ray absorption measurements at the Co and Zn
\textit{K} and \textit{L}$_{2,3}$ edges show that the Co atoms are
in a divalent state and in tetrahedral coordination, thus
substituting Zn in the wurtzite-type structure of ZnO. X-ray
magnetic circular dichroism at the Co \textit{L}$_{2,3}$ edges
reveals that the Co sublattice is paramagnetic at all temperatures
down to 2~K, both at the surface and in the bulk of the films. A
total magnetic moment of $\sim$1.2~$\mu_{\mathrm{B}}$/Co atom is
observed at 2~K and 6.5~T for Zn$_{0.9}$Co$_{0.1}$O:As, of the same
order as the paramagnetic contribution measured under similar
conditions by SQUID magnetometry, but it is reduced to
$\sim$0.55~$\mu_{\mathrm{B}}$/Co atom in Zn$_{0.75}$Co$_{0.25}$O at
5~K and 5~T. No x-ray magnetic circular dichroism signal could be
detected at the Zn \textit{L}$_{2,3}$ edges, thus excluding the
presence of a large magnetic polarization of the Zn sublattice. The
ferromagnetic component observed by bulk magnetization up to room
temperature cannot therefore be ascribed to the Co and Zn
sublattices, but it might be related to an unusual magnetic coupling
mechanism likely related to the oxygen sublattice.

The authors would like to thank B. Muller, F. Maingot, A. Derory and
B. Zada for technical assistance and R. Gusmeroli and C. Dallera for
making their multiplet calculations software available and for their
continuous assistance in performing the calculations. A. B.
acknowledges fruitful discussions with N. Hoa Hong and D. Testemale.
Work at Bessy was supported by the European Community - Research
Infrastructure Action under the FP6 "Structuring the European
Research Area" Programme (through the Integrated Infrastructure
Initiative "Integrating Activity on Synchroton and Free Electron
Laser Science" - Contract R II 3-CT-2004-506008). Part of the work
was performed at the Swiss Light Source, Paul Scherrer Institut,
Villigen, Switzerland.


\bibliography{barla}

\end{document}